\documentclass[twocolumn,npa,showpacs]{revtex4}
\usepackage{amsmath,bm}%
\usepackage{graphicx}%
\usepackage{ulem}

\begin{document}

\draft

\title{Relationship Between Structural Characters and Synchronizability of Scale-free
Networks}

\author{Jian-Guo Liu$^1$\footnote{Electric address: liujg004@yahoo.com.cn}, Yan-Zhong Dang$^1$, Qiang Guo$^2$, and Zhong-Tuo
Wang$^1$}
\address{$^1$Institute of System Engineering, Dalian University of Technology,
Dalian 116023, P R China\\ $^2$School of Science, Dalian
Nationalities University, Dalian 116600, P R China}


\begin{abstract} \textnormal{\small {Using Memory Tabu Search(MTS) algorithm,
we investigate the relationship between structural characters and
synchronizability of scale-free networks by maximizing and
minimizing the ratio $Q$ of the eigenvalues of the coupling matrix
by edge-intercrossing procedures. The numerical results indicate
that clustering coefficient $C$, maximal betweenness $B_{max}$ are
two most important factors to scale-free network synchronizability,
and assortative coefficient $r$ and average distance $D$ are the
secondary ones. Moreover, the average degree $\langle k\rangle$
affects the relationship between above structural characters and
synchronizability of scale-free networks, and the minimal $Q$
decreases when $\langle k\rangle$ increases.}}
\end{abstract}


\maketitle
Based on nonlinear dynamics, synchronization in coupled dynamical
systems has been studied for many years. It is observed in a variety
of natural, social, physical, and biological systems and has found
applications in a variety of fields
\cite{SSS1,SSS2,SSS3,SSS4,SSS5,SSS6,SSS7}. In particular,
synchronization in networks of coupled chaotic systems has received
a great deal of attention over the past two decades
\cite{SY1,SY2,SY3,SY4}. However, most of these works have been
concentrated on networks with regular topological structures such as
chains, grids, lattices, and fully connected
graphs\cite{AD1,AD2,AD3,AD4}. Recent empirical studies have
demonstrated that many real-world networks have two common
statistical characteristics: small-world effect\cite{WS98} and
scale-free property\cite{BA99}, which cannot be treated as regular
or random networks. Recently, an increasing number of studies have
been devoted to investigating synchronization phenomena in complex
networks with small-world and scale-free topologies
\cite{VV1,VV2,VV3,VV4}.

One of the goals in studying network synchronization is to
understand how the network topology affects the synchronizability.
The network synchronizability can be measured well by the eigenratio
$Q$ of the largest eigenvalue and the smallest nonzero eigenvalue
\cite{PC,Q2,ZT1,Q4}; thus, our work is to understand the
relationship between network structure and its eigenvalues. Since
there are several topological characters of scale-free networks,
what is the most important factor by which the synchronizability of
the system is mainly determined?

In this brief report, we studied the relationship between structural
characters and synchronizability of scale-free networks. Some
detailed comparisons among various networks have been done,
indicating the network synchronizability will be stronger with
smaller heterogeneity, which can be measured by the variance of
degree distribution or betweenness distribution \cite{ZT1,ZT2,ZT3},
but the strict and clear conclusions have not been achieved because
that previous studies are of both varying average distances and
degree variances. Another extensively studied one is average
distance $D$. Some works indicated the average distance $D$ is one
of the key factors to network synchronizability \cite{ZT4}. However,
we have not achieved the consistent conclusion \cite{ZT1,VV2,VV4}.
Some researchers considered that the randomicity is the more
intrinsic factors leading to better synchronizability \cite{IN},
which means that the intrinsic reason making small-world and
scale-free networks having better synchronizability than regular
ones is their random structures. Recently, several researches
examine the effect of clustering coefficient on the synchronization
by using Kuramoto model \cite{AD5} or master stability function
\cite{AD6,AD7}. Other researchers focus on the role played by
maximal betweenness $B_{max}$, they found the network
synchronizability will be better with smaller $B_{max}$
\cite{Syn4,ZT2}. Zhao {\it et. al} \cite{ZT3} enhanced the
synchronizability by structural perturbations, they found that
maximal betweenness plays a main role in network synchronization
\cite{Zhao1}. Zhou {\it et. al} \cite{ZT4} studied the average
distance $D$ to synchronizability by crossed double cycle. However,
a network contain several statistical characteristics, such as
degree distribution $P(k)$, average distance $D$, clustering
coefficient $C$, maximal betweenness $B_{max}$ and so on. In the
previous works, if one wants to show clearly how a structural
character affects the network synchronizability, such as average
distance $D$, he would investigate the network synchronizability
with different $D$ while keeping other structural characters
constant approximately. However, this method neglect the influence
made by the initial network structural characters. In fact, the
network functions, such as the synchronizability, are affected by
these characteristics simultaneously. Therefore, we should
investigate these structural factors holistically. In order to find
the real factors affect network synchronization and eliminate the
influence made by the initial networks, we maximize and minimize the
eigenratio $Q$ by MTS algorithm \cite{ETS} from the same initial
network. The structural characters which change dramatically from
maximal $Q$ to minimal $Q$ are the key factors influence network
synchronizability, while the ones change little is not.

We investigate the synchronizability of a class of continuous-time
dynamical networks with scale-free topology. Based on the
synchronization criterion, we maximize and minimize the ratio $Q$ of
the eigenvalues of the coupling matrix by edge-intercrossing
procedures, which provides a way for observing the correlation
between the synchronizability and those characteristics by keeping
the degree distribution unchanged.

We start by considering a network of $N$ linearly coupled identical
oscillators. The equation of motion reads
\begin{equation}
\dot{x}^i={\bf F}(x^i)+\sigma\sum_{j=1}^NG_{ij}{\bf H}(x^j), \ \
i=1,\cdots,N,
\end{equation}
where $\dot{{\bf x}}={\bf F}({\bf x})$ govern the local dynamics of
the vector field $x^i$ in each node, ${\bf H}({\bf x})$ is a linear
vectorial function, $\sigma$ is the coupling strength, and $G$ is a
coupling matrix.

Stability of the synchronous state ${x}^i(t)={x}^s(t), \
i=1,\cdots,N$ can be accounted for by diagonalizing the linear
stability equation, yielding $N$ blocks of the form
$\dot{\zeta}_i={\bf JF}({\bf x}_s)\zeta_i-\sigma\lambda_i{\bf
H}(\zeta_i)$, where ${\bf J}$ is the Jacobian operator. Replacing
$\sigma\lambda_i$ by $\nu$ in the equation, the master stability
function(MSF) \cite{PC} fully accounts for linear stability of the
synchronization manifold. For a large class of oscillatory systems,
the MSF is negative in a finite parameter interval
$I_{st}\equiv(\nu_1\leq \nu\leq \nu_2)$\cite{PC}. When the whole set
of eigenvalues (multiplied by $\sigma$) enters the interval
$I_{st}$, the stability condition is satisfied. This is accomplished
when $\sigma\lambda_2>\nu_2$ and $\sigma\lambda_N<\nu_2$
simultaneously. As $\nu_2$ and $\nu_1$ depend on the specific choice
of ${\bf F}({\bf x})$ and ${\bf H}({\bf x})$, the key quantity for
assessing the synchronization of a network is the eigenratio
\begin{equation}
Q=\lambda_N/\lambda_2,
\end{equation}
which only depends on the topology of the network. The small
$\lambda_N/\lambda_2$ is, the more packed the eigenvalues of $G$
are, leading to an enhanced $\sigma$ interval for which stability is
obtained\cite{Syn3}. In this paper, we will not address a particular
dynamical system, but concentrate on how the network topology
affects eigenratio $Q$.

The processes of heuristic algorithm, named MTS, is as follows.
\begin{description}
\item[Step 1.] Generate an initial matrix $G_0$ of the extensional BA
network \cite{XBA1,XBA2} with $N$ nodes and $E$ edges. Set the
optimal network' coupling matrix $G_{k}^*=G_0$ and the optimal
network of taibu table $G_{k}=G_0$, and the time step $k=0$. Compute
the ratio $Q$ of $G_k^*$.

\item[Step 2.] If a prescribed termination condition is satisfied,
stop; Otherwise intercrossing a pairs of edges chosen randomly based
on the network remains connected, denote by ${\bf G}$.

\item[Step 3.] If the ratio $Q$ of ${\bf G}$, denoted by $Q_{G}$,
satisfying $Q_{G}<Q_{G_k^*}$, $Q_{G_k^*}=Q_{G}$, else if $Q_G\leq
Q_{G_k}$, $G_{k+1}:=G$. When $Q_G>Q_{G_k}$, if $G$ does not satisfy
the tabu condition, $|Q_{G_k}-Q_G|/R_G>\delta$ (where $\delta$ is a
random number between 0.5 and 0.75), $G_{k+1}=G_k$, else
$G_{k+1}=G$. Go to Step 2.
\end{description}
Since the MTS algorithm is heuristic, it can only find the
approximate optimal solution. Thus, the termination condition of
Step 2 should confirm by the experimentation solution.

The numerical results are experimented on extensional BA model for
different network scales. The statistical properties of the optimal
networks show similar trends. After many numerical experimentations,
we set the termination condition for maximizing $Q$ as 8000 time
steps and the one for minimizing $Q$ as 3000 time steps, which can
obtain the stability value using MTS algorithm.

\begin{figure}
  \begin{center}
       \center \includegraphics[width=9.5cm]{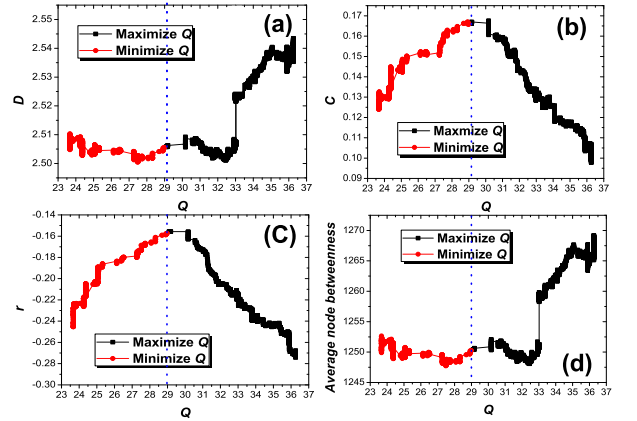}
       \caption{(Color online) The structural characters vs the eigenratio
       $Q$. (a) Average distance $D$. (b) Clustering coefficient $C$. (c) Assortative coefficient $r$.
       (d) Maximal node betweenness $B_{max}$. The blue dot line
       denotes the state of the initial network.  The data are averaged over ten
       independent runs of network size $N=500$.}\label{F001}
 \end{center}
\end{figure}

\begin{figure}
  \begin{center}
       \center \includegraphics[width=9.5cm]{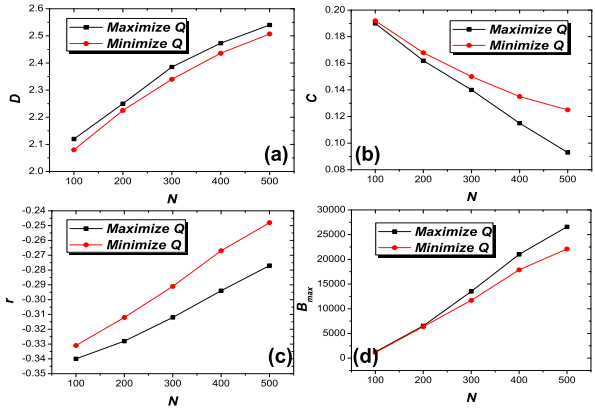}
       \caption{(Color online) $B_{max}$
       obtained by maximizing and minimizing $Q$ vs network size
       $N$. }\label{F002}
 \end{center}
\end{figure}


\begin{figure}
  \begin{center}
       \center \includegraphics[width=9.5cm]{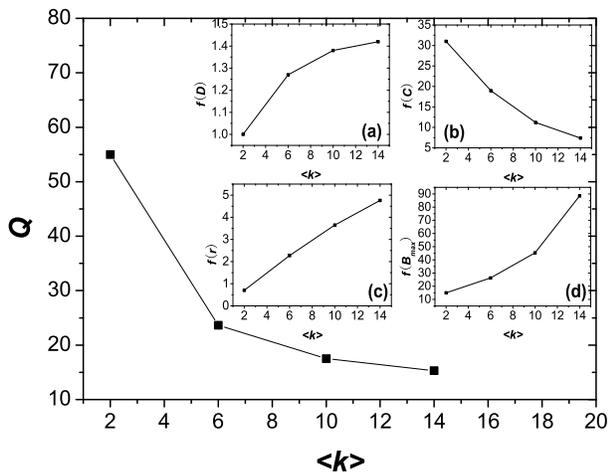}
       \caption{(Color online) Relationship between $Q$ and
       average degree $\langle k\rangle$. The inset shows the functions of the structural
characters to different $\langle k\rangle$.}\label{F004}
 \end{center}
\end{figure}

We start from a network of size $N=100$, 200, 300, 400, 500 and the
average degree $\langle k\rangle=6$ and then perform the
optimization precesses. At each time step, we record the structural
properties, such as $D$, $C$, $r$ and average node betweenness, when
the objective function $Q$ is reduced. Let $D^{min}$, $C^{min}$
$r^{min}$ and $B_{max}^{min}$ denote the stable value when $Q$
reaches its minimal value $Q_{min}$, and $D^{max}$, $C^{max}$,
$r^{max}$ and $B^{max}_{max}$ denote the stable value when $Q$
reaches its maximal value $Q_{max}$. Define the {\it relative
diversity function} of the structural character $x$ as follows
\begin{equation}
f(x)=\frac{|x^{max}-x^{min}|}{x^{min}}*100,
\end{equation}
which can denote the difference of structural character $x$ to
$Q_{max}$ and $Q_{min}$. The larger $f(x)$ is, the structural
character $x$ change dramatically when the network leave far from
its optimal synchronizability state, which means $x$ is more
relevant to network synchronizability.

Figure \ref{F001}. (a) demonstrates that $D$ remains stable when
minimizing $Q$, while increase a little when maximizing $Q$. Figure
\ref{F001}. (b), (c) show $C$ and $r$ decrease to a stable value
when maximizing and minimizing $Q$, and $C^{max}$ and $r^{max}$ are
both smaller than $C^{min}$ and $r^{min}$. The difference between
$C^{min}$ and $C^{max}$, $r^{min}$ and $r^{max}$ means that the two
structural characters are relevant to synchronizability of
scale-free networks. Figure \ref{F001}. (d) gives the change trend
of average node betweenness when maximizing and minimizing $Q$,
which is consistent with $D$. Figure \ref{F002}. (a)-(d) demonstrate
the stable value of $D$, $C$, $r$ and $B_{max}$ when $N=100, 200,
300, 400, 500$. From Fig.\ref{F002}, one can obtain that when
$N=500$, $f(D)=1.27$, $f(C)=20.97$ and $f(r)=1.27$ and
$f(B_{max})=19.04$. Moreover, one can see that the relative
diversity of $C$, $r$ and $B_{max}$ become large, while the one of
$D$ remain constant, which indicates that the influence produced by
the structural characters $C$, $r$ and $B_{max}$ to
synchronizability of scale-free networks would become great when $N$
become large. Furthermore, we investigate the relationship between
average degree $\langle k\rangle$ and $f(x)$. Figure \ref{F004}
demonstrats the $Q_{min}$ obtained by MTS algorithm to different
$\langle k\rangle$ when $N=500$. The inset gives the functions of
the structural characters obtained by different $\langle k\rangle$.
From the inset, one can see that if $\langle k\rangle$ increase, the
function of $D$, $r$ and $B_{max}$ increases while the function of
$C$ decreases, which means that the influence of the structural
characters to network synchronizability is affected by $\langle
k\rangle$. When $\langle k\rangle$ increases, $B_{max}$, $D$ and $r$
become more relevant to synchronizability of scale-free networks,
while $C$ become less relevant.

In summary, using the MTS optimal algorithm, we maximized and
minimized the network synchronizability by changing the connection
pattern between different pairs of nodes while keeping the degree
variance unchanged. Starting from extensional BA networks, we found
the relationship between structural characters and synchronizability
of scale-free networks. The numerical results indicate that $D$,
$C$, $r$ and $B_{max}$ influence network synchronizability
simultaneously. Especially, $C$ and $B_{max}$ are the two most
important structural characters which affect synchronizability of
scale-free networks, assortative coefficient $r$ is the secondary
character and $D$ is the last one. Furthermore, the relationship is
affected by the average degree $\langle k\rangle$, and the maximal
synchronizability of scale-free networks increases when $\langle
k\rangle$ increases.

The authors thank W. -X. Wang, T. Zhou and Bing Wang for their
valuable comments and warm discussions. This work has been partly
supported by the Natural Science Foundation of China under Grant
Nos. 70431001 and 70271046.

\end{document}